\documentclass[a4paper,11pt]{article}
\usepackage{pos}
\usepackage{graphicx}
\graphicspath{ {./images/} }
\usepackage{amsmath}
\begin{document}
\title{Lee-Yang edge singularities in 2+1 flavor QCD with imaginary chemical potential}

\author*[a]{Guido Nicotra}
\author[b]{Petros Dimopoulos}
\author[b]{Lorenzo Dini}
\author[b]{Francesco Di Renzo}
\author[a]{Jishnu Goswami}
\author[a]{Christian Schmidt}
\author[b]{Simran Singh}
\author[b]{Kevin Zambello}
\author[a]{Felix Ziesche}

\affiliation[a]{Fakultät für Physik, Universität Bielefeld, D-33615 Bielefeld, Germany;}
\affiliation[b]{Dipartimento di Scienze Matematiche, Fisiche e Informatiche, Università di Parma, Parma, Italy}

\emailAdd{gnicotra@physik.uni-bielefeld.de}
\emailAdd{petros.dimopoulos@unipr.it}
\emailAdd{lorenzo@physik.uni-bielefeld.de}
\emailAdd{francesco.direnzo@unipr.it}
\emailAdd{jishnu@physik.uni-bielefeld.de}
\emailAdd{schmidt@physik.uni-bielefeld.de}
\emailAdd{simran.singh@unipr.it}
\emailAdd{kevin.zambello@studenti.unipr.it}
\emailAdd{fziesche@physik.uni-bielefeld.de}


\abstract{We present results of the location of the closest singularities in the complex chemical potential plane using a novel method. These results are obtained with (2+1)-flavor of highly improved staggered quarks (HISQ) on lattices with temporal extent of $N_\tau=4,6$. We show that the scaling is consistent with the expected scaling of the Lee-Yang edge singularities in the vicinity of the Roberge-Weiss (RW) transition. We determine various non-universal parameters using 3D Ising model scaling functions that map QCD in the scaling region of the RW transition. Furthermore, as a preliminary result we discuss how the Lee-Yang edge singularity can be used to probe the chiral phase transition in QCD. The singularity obtained close to the chiral phase transition temperature $T_c$ seems to be in agreement with the expected scaling of the Lee-Yang edge singularity.
As an outlook, we discuss the scaling of the Lee-Yang edge singularity in the vicinity of a possible critical end point in QCD, at even lower temperatures. In the future, such a scaling analysis might hint on the existence and the location of the critical end point.\\
The work presented here is a part of an ongoing project of Bielefeld Parma joint collaboration.}

\FullConference{The 38th International Symposium on Lattice Field Theory\\
Zoom/Gather@MIT\\
July 26-30 2021
}


\maketitle


\section{Introduction}
During the last decades lattice QCD has proven to be a very powerful tool in the study of the QCD phase diagram, providing new results and a deeper understanding of the properties of the strongly interacting matter at finite temperature and vanishing chemical potential~\cite{review}.
Recent studies \cite{Goswami:2020yez,Bollweg:2021vqf} show that the confined hadronic matter changes its degrees of freedom close to the QCD crossover temperature in the case of a vanishing chemical potential \cite{Aoki:2006we}. It has been conjectured that this analytic crossover turns into a first order phase transition in the case of large baryon chemical potential. Following these, one expects that between these two lines exists a second order Z(2) critical end point.\\
However, the sign problem invalidates the importance sampling techniques when one introduces a non-zero baryon chemical potential in the lattice action. Several different methods were developed to circumvent the numerical sign problem. For example, the Taylor expansion is a powerful method which allow us to extract meaningful information from small chemical potential regions \cite{Gavai:2008zr, Allton:2005gk}. 
Another possibility consists in the introduction of a imaginary chemical potential which doesn't produce any complex phase from the fermion determinant, allowing direct calculations on the lattice \cite{deForcrand:2003bz, DElia:2002tig} .\\
In our work we have performed calculations with an imaginary baryon chemical potential and we have used the Taylor series to build Padé approximants. The details of our new method were already presented in this conference \cite{SimProc}. This allowed us to study their zeroes and poles. The original idea from Lee, Yang and Fischer \cite{Yang:1952be, Fisher:1978pf} is that one can extract meaningful information about different phase transitions from stable singularities.

\section{Lee-Yang Edge Singularities on the lattice}

\textit{Lee-Yang edge singularities} (LYEs) corresponds to a second order phase transition for a finite symmetry breaking field. A second order phase transition is identified from the divergence of the correlation length in the vicinity of the critical point. This means that the fluctuations, namely the susceptibilities, become correlated over all distances and the entire system is forced to be a unique critical phase. The behavior of critical properties in the vicinity of a second order phase transitions can be described by a limited number of universality classes defined by the \textit{fundamental symmetry} of the system. In particular, different systems belonging to the same universality class have the same critical exponents which characterize the behavior of physical quantities near the continuous phase transitions.\\
The contributions to the order parameter $M$ can be written as the sum of a regular and singular part \cite{Ejiri:2009ac}:
\begin{equation}
    M(t,h)=h^\frac{1}{\delta}f_G(z)+M_{reg}(t,h)
    \label{eq:mag_state}
\end{equation}
where $z=t/|h|^{1/\beta\delta}$ is the scaling variable with $\beta$ and $\delta$ critical exponents, $t$ is the reduced temperature and $h$ is the symmetry breaking field. Eq.(\ref{eq:mag_state}) is known as the Magnetic Equation of state. Prolonging $h$ in the complex plane, the \textit{scaling function} $f_G$ shows a universal singularity at $z=z_c$ which is called LYEs. One has now to properly define the variables $t$ and $h$ accordingly to the symmetry group of the transition, the critical point of that transition is then found in the limits $h\rightarrow 0$ and $t\rightarrow 0$.\\
In this proceeding we discuss three different second order transitions by studying the LYEs: the RW transition at imaginary chemical potential, the chiral transition at vanishing chemical potential and the QCD critical endpoint at real chemical potential. We used the Padé approximant methods (details on our method in \cite{SimProc,Dimopoulos:2021vrk}) to extract the singularities. In the following sections we determine LYEs for RW and chiral transition. We also determine the value of the non-universal constant $z_0$ from relevant scaling functions.

\begin{figure}[h]
    \centering
	\includegraphics[scale=0.58]{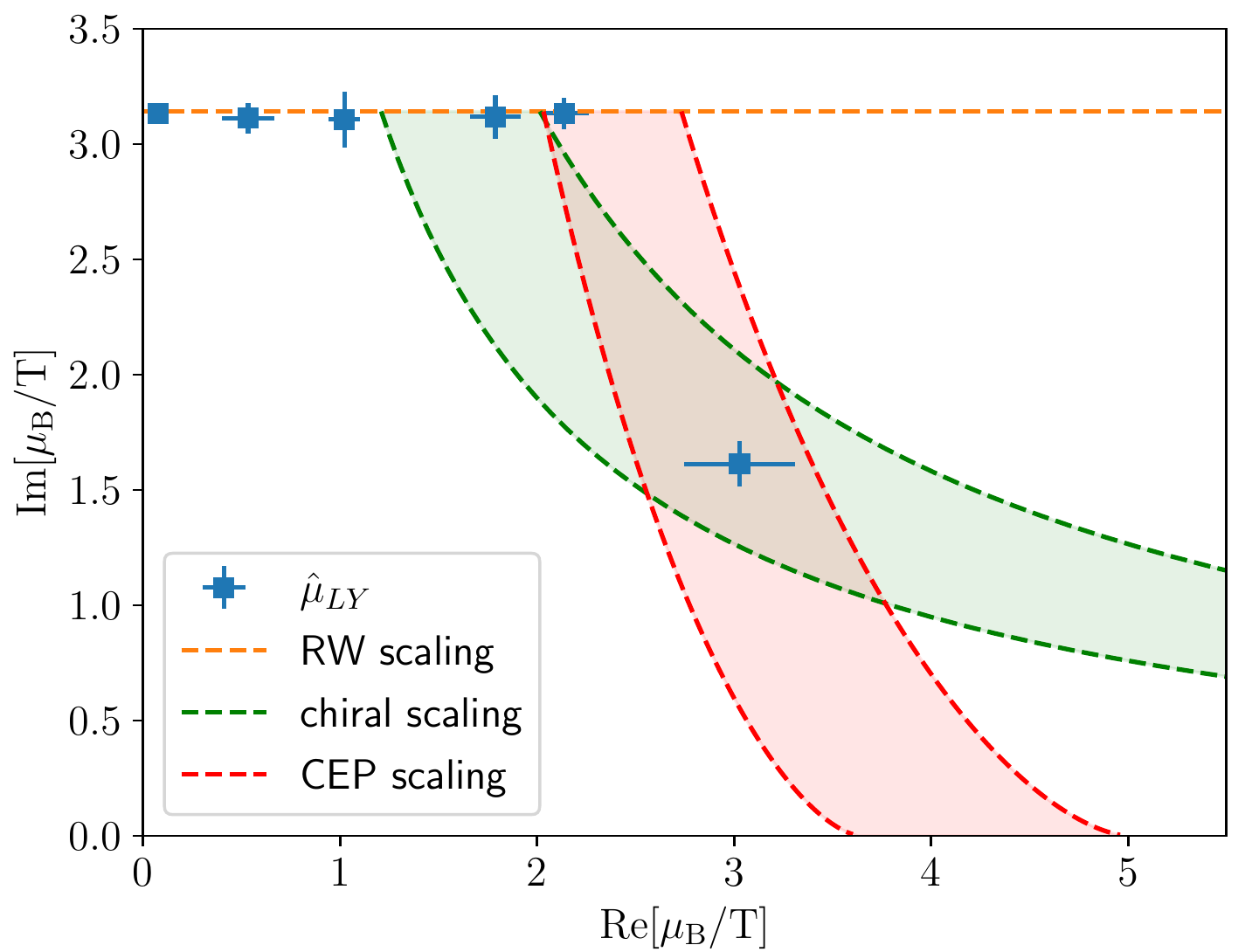}
	\caption{An overview of all the scaling functions we analyzed. The yellow line is referred to the Z(2) RW transition scaling, the green band to the O(2) chiral transition scaling and the red band to the Z(2) critical endpoint.}
	\label{GenScaling}
\end{figure}

\subsection{LYEs for RW transition}
The partition function of QCD with imaginary chemical potential has a periodicity of $2\pi$ in the baryon chemical potential ($\mu_B)$ and at $\hat{\mu}_B=\mu_B/T=i\pi$ one can observe the RW transition. In (2+1)-flavor QCD with physical value of quark masses with HISQ action , the symmetry associated to the transition is Z(2) \cite{JishnuPoS,Goswami:2019exb} and the order parameter 
is the number density $\chi_1^B$\footnote{The general definition for susceptibilities we used is: 
\begin{equation*}
    \chi_{ijk}^{BQS}=\frac{1}{VT^3}\frac{\partial^{i+j+k}}{\partial(\mu_B/T)^i\partial(\mu_Q/T)^j\partial(\mu_S/T)^k}lnZ
\end{equation*} where $Z$ is the partition function, $V$ the volume of the system \cite{Allton:2005gk}.}. The scaling fields are defined as follows:
\begin{align*}
    h=\frac{1}{h_0}\frac{\hat{\mu}_B-i\pi}{i\pi}&
    &t=\frac{1}{t_0}\frac{T-T_{RW}}{T_{RW}}\,,
\end{align*}
after solving the equation $z=z_c$, one finds the temperature scaling for the LYEs:
\begin{equation}
    Re\Big(\frac{\mu_B}{T}\Big)=\pm\pi\Big(\frac{z_0}{|z_c|}\Big)^{\beta\delta}\Big(\frac{T_{RW}-T}{T_{RW}}\Big)^{\beta\delta}
    \label{temp_scaling1}
\end{equation}
\begin{equation}
    Im\Big(\frac{\mu_B}{T}\Big)=\pm\pi
    \label{temp_scaling2}
\end{equation}
Using Eq.(\ref{temp_scaling1}) and Eq.(\ref{temp_scaling2}) we were able to extract the value of the normalization constant $z_0$, defined as $z_0=h_0^{1/\beta\delta}/t_0$,  performing a scaling fit of the LYEs.\\
We also used the Magnetic Equation of State in order to evaluate $z_0$, isolating $f_G$ from  Eq.(\ref{eq:mag_state}):
\begin{equation}
    f_G(z)=\frac{M(t,h)-M_{reg}(t,h)}{h^{1/\delta}}
    \label{inverted}
\end{equation}

\begin{figure*}[h!] 
        \centering
    	\includegraphics[scale=0.27]{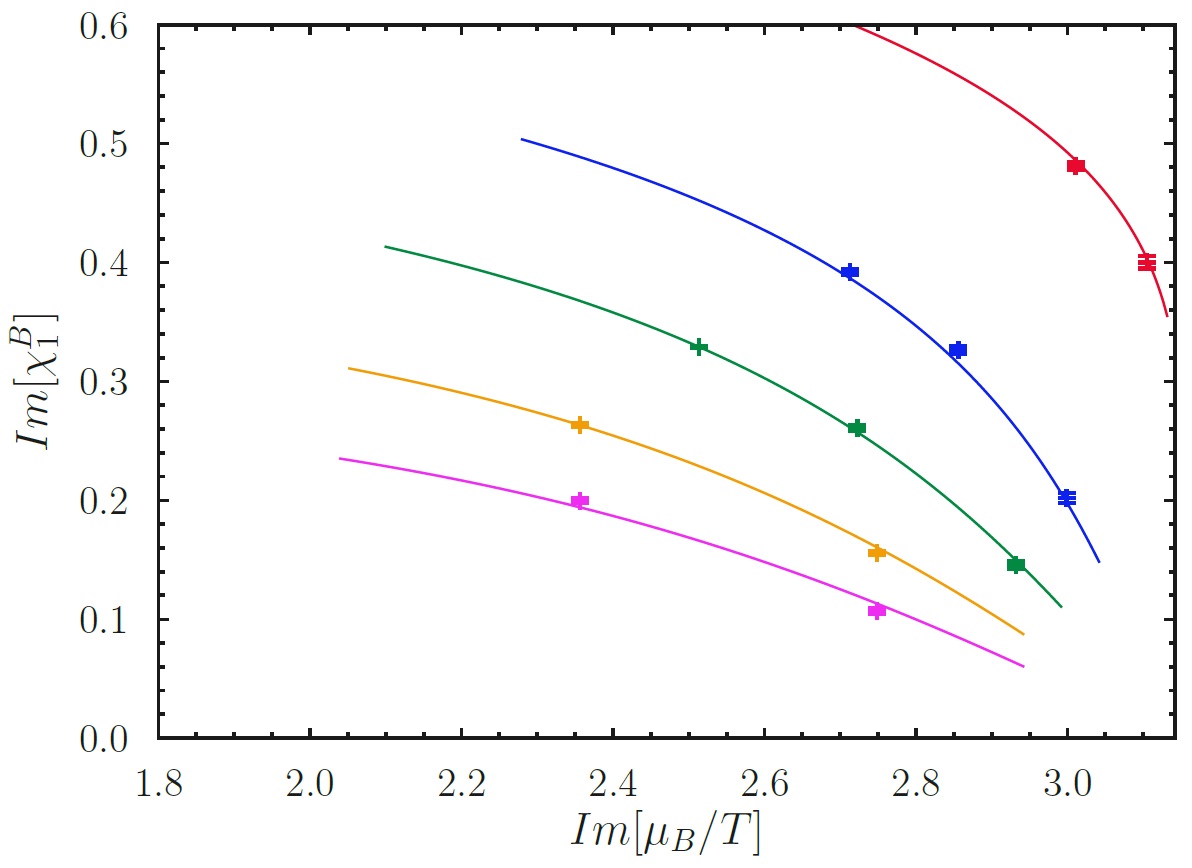}
        \hskip1cm
	    \includegraphics[scale=0.27]{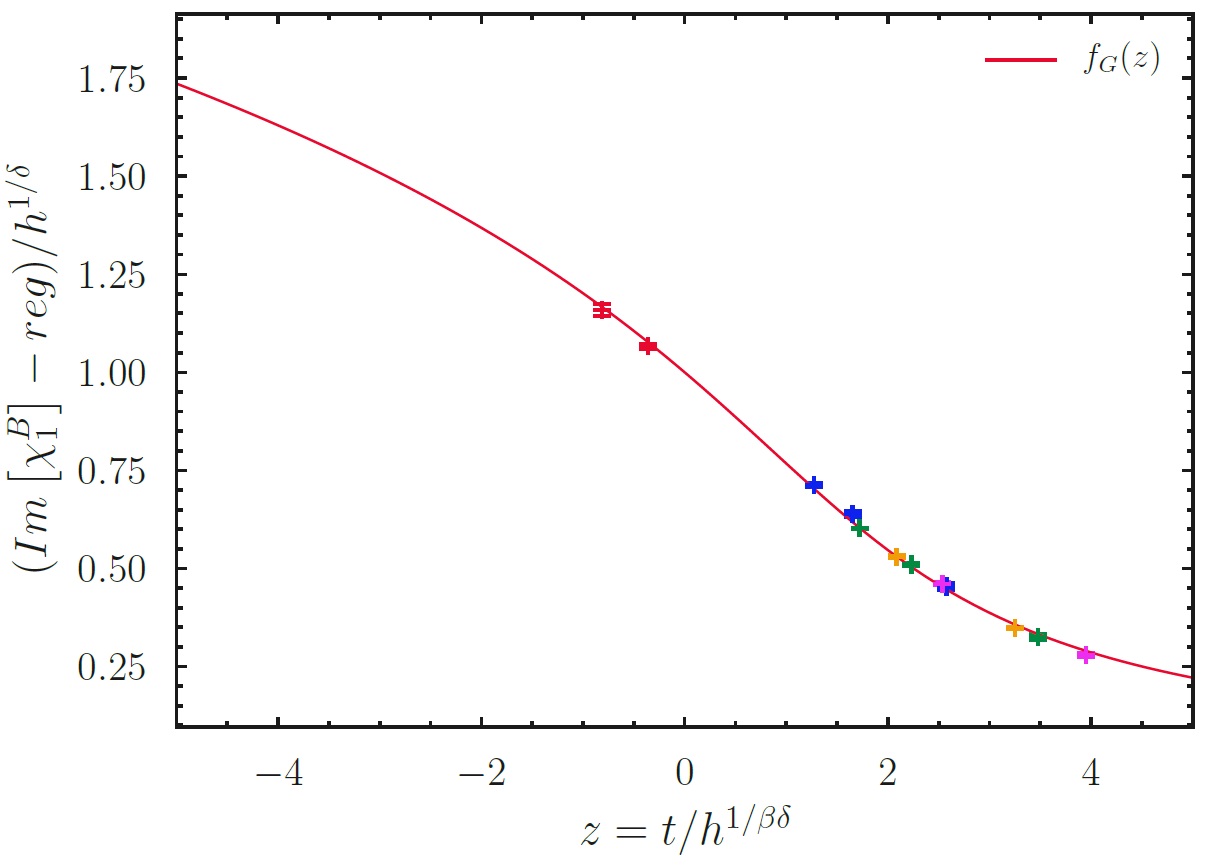}
	    \caption{The first order susceptibility at different temperatures (left) and the behavior of the scaling function (right). It is possible to see how all the points on the first figure are mapped along the scaling function line.}
         \label{fig:appendixB_Froiss}
\end{figure*}

\subsection{LYEs for chiral transition}
In the case of a (2+1)-flavor QCD theory with staggered fermions and away from the continuum limit, one expects that the universal scaling of the chiral transition belongs to the universality class of the 3D $O(2)$ model. We have set the scaling variables as:
\begin{align*}
    h=\frac{1}{h_0}\Big(\frac{m_l}{m_s}\Big)&
    &t=\frac{1}{t_0}\Big[\frac{T-T_c}{T_c}-\kappa_2^B\Big(\frac{\mu_B}{T}\Big)^2\Big]\,,
\end{align*}
where the ratio $m_l/m_s$, the mass of the light and strange quarks, is the order parameter of the transition and $\kappa_2^B$ \cite{HotQCD:2018pds}, the curvature parameter along the line of $T_c(\mu_B)$. From these we obtain the subsequent scaling \cite{Mukherjee:2019eou}:
\begin{equation}
    \frac{\mu_B}{T}=\Big[\frac{1}{\kappa_2^B}\Big(\frac{T-T_c}{T_c}-\frac{z_c}{z_0}\Big(\frac{m_l}{m_s}\Big)^{1/\beta\delta}\Big)\Big]^{1/2}
\end{equation}

\subsection{Lattice setup}
We performed simulations with Highly Improved Staggered Quarks in the (2+1)-flavor QCD with Rational Hybrid Monte Carlo (RHMC) algorithm. The calculations has been done with lattice sizes $24^3\times 4$ and $36^3\times 6$ at different imaginary chemical potentials. The conversion of beta to temperature values in MeV at finite lattice spacing has been done using the parametrization given in \cite{Bollweg:2021vqf,HotQCD:2014kol}. For more details on the choice of parameters please see~\cite{Dimopoulos:2021vrk}.\\

\section{Results}
\subsection{RW transition}
To study the RW transition we used the lattice size $N_\sigma=24^3\times N_\tau=4$ which allowed us to take advantage of the value of RW critical temperature $T_{RW}=201$ MeV already measured in \cite{JishnuPoS,Goswami:2019exb}. In Fig.(\ref{fig:appendixB_Froiss}) on the left we show the behavior of $\chi_1^B$ measured at different temperatures. 

\begin{figure*}[h!] 
        \centering
    	\includegraphics[scale=0.35]{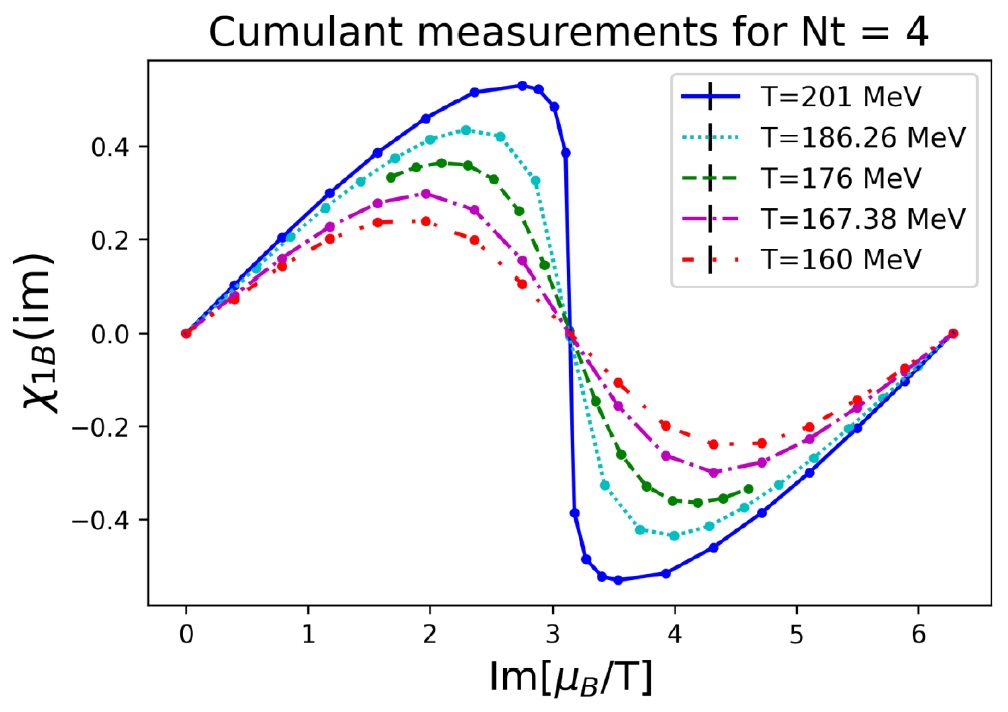}
        \hskip1cm
	    \includegraphics[scale=0.4]{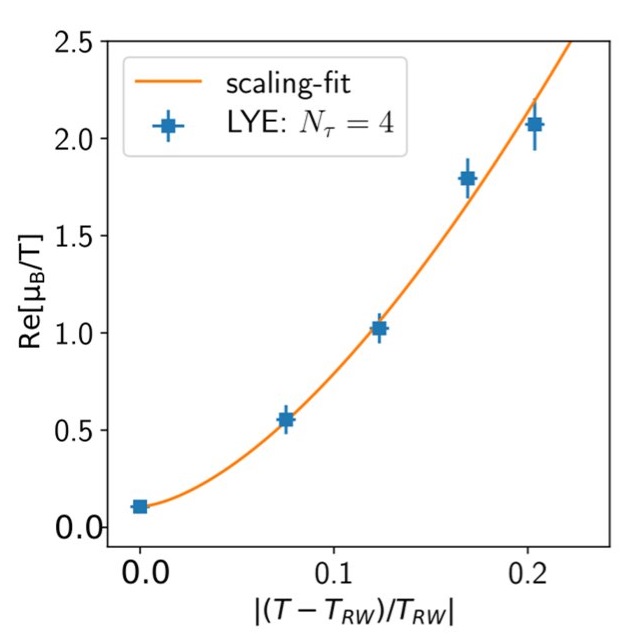}
	    \caption{$\chi_1^B$ at different temperatures (left) and the temperature scaling fit of the LYEs (right).}
         \label{fig:appendixB_Froiss}
\end{figure*}
In Fig.(\ref{fig:appendixB_Froiss}) on the right we plotted the behavior of LYEs measured at different temperatures. We performed a fit of the real parts of the singularities using the function:
\begin{equation}
    f(x)=a\cdot t^{\beta\delta}+b
\end{equation}
The imaginary parts are found to be consistent with $\hat{\mu}_B=i\cdot\pi$. We used the well known critical exponents from the 3D Ising model $\beta=0.3264$ and $\delta=4.789$, we have taken $z_c=2.452\pm 0.025$ from Ref.\cite{Connelly:2020gwa}. Our preliminary result for the normalization constant is $z_0=9.28\pm 0.26$ and $\chi^2_{test}\approx 0.8$. 
It's possible to see that the real part of the LYEs at $T_{RW}$ doesn't vanish entirely as expected, which is related to finite volume effects still to fully understand.\\
We compared the obtained result of $z_0$ with the one calculated using the Magnetic Equation of State. We have performed a fit of Eq.(\ref{inverted}) with the function:
\begin{equation}
    M_{reg}(t,h)=a_2\cdot t\cdot h
\end{equation}
finding the value $z_0=5.316\pm 0.017$ with a $\chi^2_{test}\approx 6$. The discrepancy of the two values is quite large and it might be due to several factors. 
First of all, a larger number of fitting parameters; secondly, since the scaling fit is not performed at constant $z$, extra fits of the independent normalization parameters $t_0$ and $h_0$ are necessary; for last, a well balanced choice of the distance from the transition is needed: as closer to the phase transition one takes measures as larger is the variation of $z$, while, at large distances from the transition, the corrections originated from the regular contribution $M_{reg}$ increase.

\newpage
\subsection{Chiral transition}
\begin{figure}[h]
    \centering
	\includegraphics[scale=0.6]{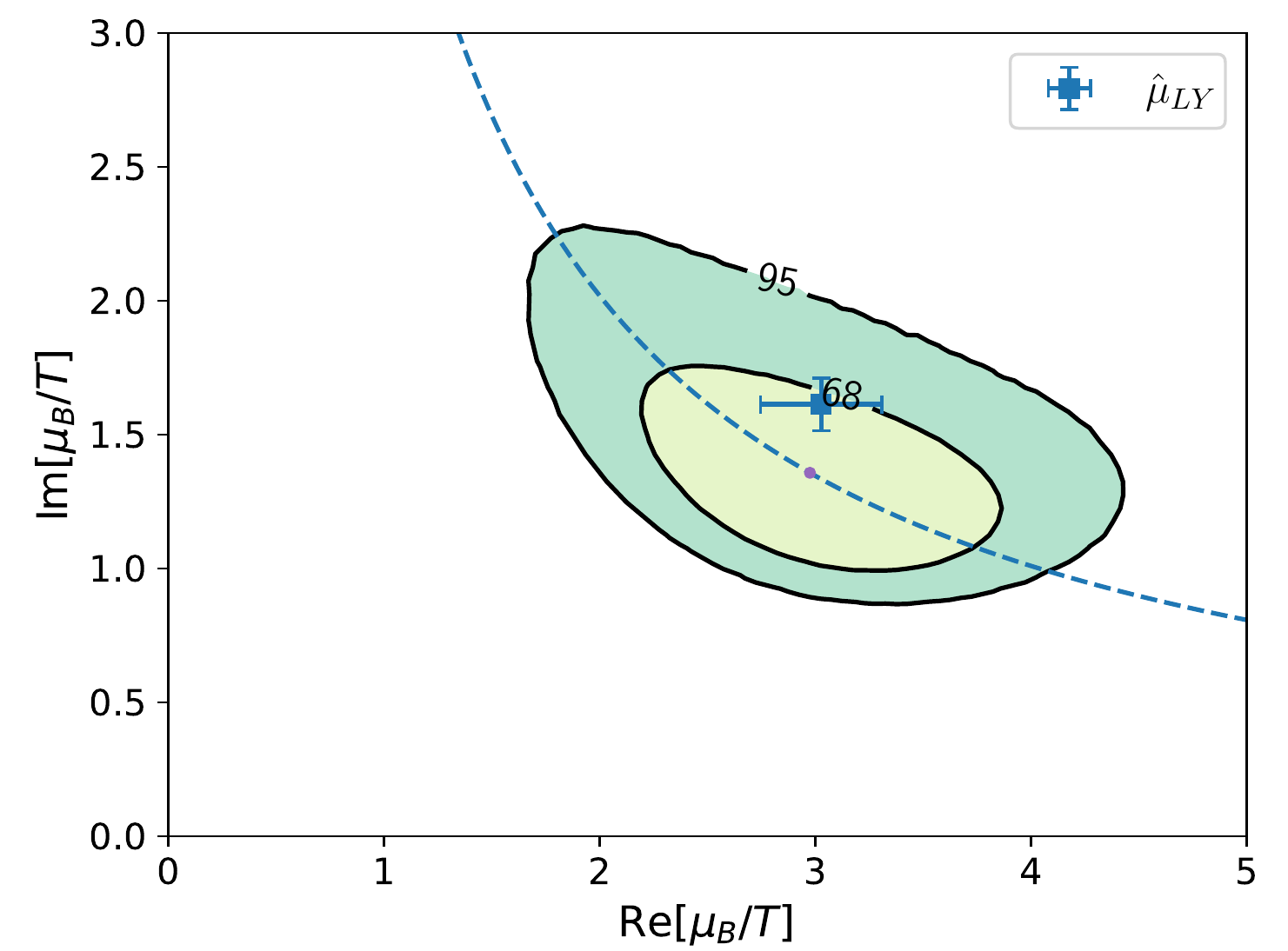}
	\caption{Comparison between the LYEs we measured and the temperature scaling prediction from HotQCD (the dashed line). The little point in the center of the green region is the predicted value at the temperature $T=145$ MeV. We also show the 68\% and 95\% confidence level.}
	\label{Chiral}
\end{figure}

Regarding the chiral  transition we performed calculations with a lattice size $N_\sigma=36^3\times 6$, physical mass ratio $m_l/m_s=1/27$ and we chose the critical temperature $T_c=147$ MeV, $\kappa_2^B=0.012\pm 0.002$, the non-universal constant $z_0=2.35$ \cite{HotQCD} and $|z_c|=2.032$ \cite{Connelly:2020gwa}.
\\
In Fig.(\ref{Chiral}) we show our LYEs and the predicted temperature dependence which was plotted using a parameter free prediction based on non-universal parameters obtained by the HotQCD collaboration. It is possible to observe that the LYEs is less than one sigma distant from the predicted value.

\subsection{Future perspective: possible location of the critical end point}
As an outlook we discuss here a possible scenario for predicting the QCD critical end point (\textbf{cep}) using our technique. 
In this case, the mapping to the universal theory with Z(2)-symmetry is actually unknown. One possibility is to use a linear map for the scaling variables:
\begin{equation*}
    h=\alpha_h(T-T_{cep})+\beta_h(\mu_B-\mu_B^{cep}),
\end{equation*}
\begin{equation*}
    t=\alpha_t(T-T_{cep})+\beta_t(\mu_B-\mu_B^{cep}).
\end{equation*}

This choice leads us to the following scaling relation \cite{Stephanov:2006dn}:
\begin{equation}
    \mu_{LY}(T)=\mu_{\text{cep}} - c_1(T-T_{cep})+ ic_2|z_c|^{-\beta\delta}(T-T_{cep})^{\beta\delta},
    \label{EPscal}
\end{equation}
where $c_1$ is the slope of the transition line at the critical point and $c_2\cdot | z_c|^{-\beta\delta}$ is a parameter depending on the relative angle between the $h$ and $t$ axes. In \cite{Basar:2021hdf} it has been shown how the parameters $c_1$, $c_2$, $T_{cep}$ and $\mu_{cep}$ can be determined in the case of the Gross-Neveu model performing a fit of Eq.(\ref{EPscal}).\\
To illustrate a possible behavior of the temperature dependence of LYEs, namely the red band in Fig.(\ref{GenScaling}), we have set the parameters as:
\begin{eqnarray*}
     c_1=0.024,&    &|z_c|^{-\beta\delta}=0.5,\\ 
 T_{cep}&=&T_{pc}(1-\kappa_2^B(\mu_{cep}/T_{pc})^2),\nonumber \\
 \text{where~}T_{pc}&=&156.5 \textrm{ MeV}.
\end{eqnarray*}
    

\section{Summary and conclusions}
We presented a new study of the universal behavior observed close to the QCD phase transitions using the LYEs and performing calculations in (2+1)-flavors lattice QCD at different imaginary chemical potential. We tested the validity of our method using the Roberge-Weiss transition. Through the $\chi^2$ analysis we found a good agreement between the temperature scaling of the measured LYEs and the theoretically predicted power law behavior. We found two different values for $z_0$ using the LYEs scaling and the Magnetic equation of state fit. The discrepancy between the two results requires further investigations. We studied a single LYEs related to the chiral transition. We compared this singularity with a parameter free prediction of the temperature dependence of the LYEs. The measured and the predicted LYEs are less than one sigma apart. At last we shortly described how our method could be applied in the search of the QCD critical end point. We reported a possible choice for the scaling variables and we have shown an hypothetical behavior of the temperature scaling. For future developments, aside from the search of the critical end point, we are interested into increasing the number of LYEs in the chiral transition region and to perform new calculations with smaller lattice spacing.

\section{Acknowledgements}
This work was supported by (i) the European Union’s Horizon 2020 research and innovation program under the Marie Skłodowska-Curie grant agreement No H2020-MSCAITN-2018-813942 (EuroPLEx), (ii) The Deutsche Forschungsgemeinschaft (DFG, German Research Foundation) - Project number 315477589-TRR 211 and (iii) I.N.F.N. under the research project {\em i.s.} QCDLAT.
This research used computing resources made available through: (i) Gauss Centre for Supercomputing on the Juwels GPU nodes at the  Jülich Supercomputing Centre (ii) Bielefeld University on the Bielefeld GPU-Cluster (iii), CINECA on Marconi100 under both the I.N.F.N.-CINECA agreement and the ISCRA C program (HP10CWD9YA project) and (v) University of Parma on the UNIPR HPC facility.

\end{document}